\documentclass[fleqn,usenatbib]{mnras}

\newcommand{\logg}{\mbox{$\rm log \it{g}$}}
\newcommand{\numax}{\mbox{$\nu_{\rm max}$}}
\newcommand{\Dnu}{\mbox{$\Delta \nu$}}
\newcommand{\muHz}{\mbox{$\mu$Hz}}
\newcommand{\teff}{\mbox{$T_{\rm eff}$}}
\newcommand{\kep}{\mbox{\em Kepler}}

\newcommand{\tess}{\mbox{$TESS$}}
\newcommand{\plato}{\mbox{$PLATO$}}
\newcommand{\amp}{\mbox{$A_{\lambda}$}}
\newcommand{\vosc}{\mbox{$v_{\rm osc}$}}
\newcommand{\sigmaphot}{\mbox{$\sigma_{\rm rms,\ \rm phot}$}}
\newcommand{\sigmarv}{\mbox{$\sigma_{\rm rms,\ \rm RV}$}}
\newcommand{\redbf}[1]{#1}

\usepackage[T1]{fontenc}
\usepackage{ae,aecompl}
\usepackage{natbib}
\usepackage{newtxtext,newtxmath}

\usepackage{graphicx}	
\usepackage{amsmath}	
\usepackage{amssymb}	

\title[Estimating RV jitter in terms of fundamental stellar properties]{Predicting radial-velocity jitter induced by stellar oscillations based on \kep\ data}

\author[Jie Yu et al.]{
Jie Yu,$^{1,2}$\thanks{E-mail: jiyu9229@uni.sydney.edu.au (JY)}
Daniel Huber,$^{3,1,4,2}$
Timothy R. Bedding$^{1,2}$
and Dennis Stello$^{5,1,2}$
\\
$^{1}$Sydney Institute for Astronomy (SIfA), School of Physics, University of 
Sydney, NSW 2006, Australia\\
$^{2}$Stellar Astrophysics Centre, Department of Physics and Astronomy, Aarhus University, Ny Munkegade 120, DK-8000 Aarhus C, Denmark\\
$^{3}$Institute for Astronomy, University of Hawai`i, 2680 Wood-lawn Drive, Honolulu, HI 96822, USA\\
$^{4}$SETI Institute, 189 Bernardo Avenue, Mountain View, CA 94043, USA\\
$^{5}$School of Physics, University of New South Wales, NSW 2052, Australia\\
}

\date{Accepted XXX. Received YYY; in original form ZZZ}

\pubyear{2015}

\begin{document}
\label{firstpage}
\pagerange{\pageref{firstpage}--\pageref{lastpage}}
\maketitle

\begin{abstract}
Radial-velocity jitter due to intrinsic stellar variability introduces challenges when characterizing exoplanet systems, particularly when studying small (sub-Neptune-sized) planets orbiting solar-type stars. In this Letter we \redbf{predicted} for dwarfs and giants the jitter due to stellar oscillations, which in velocity have much larger amplitudes than noise introduced by granulation. We \redbf{then fitted} the jitter in terms of the following sets of stellar parameters: (1) Luminosity, mass, and effective temperature:  the fit returns precisions \redbf{(i.e., standard deviations of fractional residuals)} of 17.9\% and 27.1\% for dwarfs and giants, respectively. (2) Luminosity, effective temperature, and surface gravity: The precisions are the same as using the previous parameter set. (3) Surface gravity and effective temperature:  we obtain a precision of 22.6\% for dwarfs and 27.1\% for giants. (4): Luminosity and effective temperature: the precision is 47.8\% for dwarfs and 27.5\% for giants. Our method will be valuable for anticipating the radial-velocity stellar noise level of exoplanet host stars to be found by the \tess\ and \plato\ space missions, and thus can be useful for their follow-up spectroscopic observations. We provide publicly available code (\url{https://github.com/Jieyu126/Jitter}) to set a prior for the jitter term as a component when modeling the Keplerian orbits of the exoplanets.
\end{abstract}




\section{Introduction}
The radial velocity (RV) technique has been widely used to discover exoplanets and to confirm exoplanets detected in transit surveys \citep[see][for recent reviews]{fischer16, wright17}. However, RV jitter from the host stars leads to challenges, particularly, when studying the exoplanetary signals of small (sub-Neptune-sized) planets that are expected to be detected by space-based transit missions such as \tess\ \citep{ricker14} and \plato\ \citep{rauer14}. Several methods have been developed to mitigate effects of stellar RV jitter, including the de-correlation magnetic activity indices \citep{saar98,isaacson10}, time-averaging of rapid oscillations \citep{Dumusque11}, and modeling correlated stellar noise using Gaussian Processes \citep{haywood14,rajpaul15} including simultaneous photometric observations \citep{Grunblatt15,Giguere16}. However, as of yet there are only few quantitative tools to predict the expected level of RV jitter for a given star, which is critical to planning and prioritizing spectroscopic follow-up observations of transiting planets. 

The RV jitter mainly comes from four sources: stellar oscillations, granulation (super-granulation), short-term activity from stellar rotation, and long-term activity caused by magnetic cycles \citep[see][and references therein]{dumusque16, dumusque17}. For dwarfs, the oscillations and granulation have timescales \redbf{on the order of} minutes, while the short- and long-term activity has a longer timescale, typically greater than tens of days. In this study, we will quantify the short-timescale jitter caused by the stellar oscillations in terms of fundamental stellar properties for a wide range of evolutionary states. We emphasize that, unlike in photometry, granulation in velocity has much lower amplitude than the oscillations \citep{bedding06}, and hence the results presented here can be used to predict RV jitter over a wide range of stars.

Relatively few stars so far have RV data with sufficient cadence to do seismology, so it is difficult to calibrate a RV jitter scaling relation as a function of stellar parameters. Fortunately, analysis of photometric time series can shed light on the RV jitter \redbf{\citep{aigrain12,bastien14}}. The \kep\ photometric time series \redbf{have} been widely explored to study the stellar oscillations in dwarfs and giants \citep[see a review by][]{chaplin13b}. \citet{KB95} proposed that the spectroscopic and photometric oscillation amplitudes are convertible between each other. Moreover, it has been widely demonstrated that asteroseismology is able to provide accurate estimates of stellar parameters, based on photometric data sets \redbf{\citep[see][for reviews]{chaplin13b, hekker17}}. These facts suggest that asteroseismic analyses on the photometric time series allow us to estimate the RV jitter in terms of stellar parameters.

In this Letter, we provide simple relations to predict the RV jitter from stellar parameters, luminosity, mass, effective temperature, and surface gravity. We \redbf{also} provide public code for implementing \redbf{these} predictions.

\section{Method and Data}
\label{method}
\begin{figure}
\begin{center}
\resizebox{\columnwidth}{!}{\includegraphics{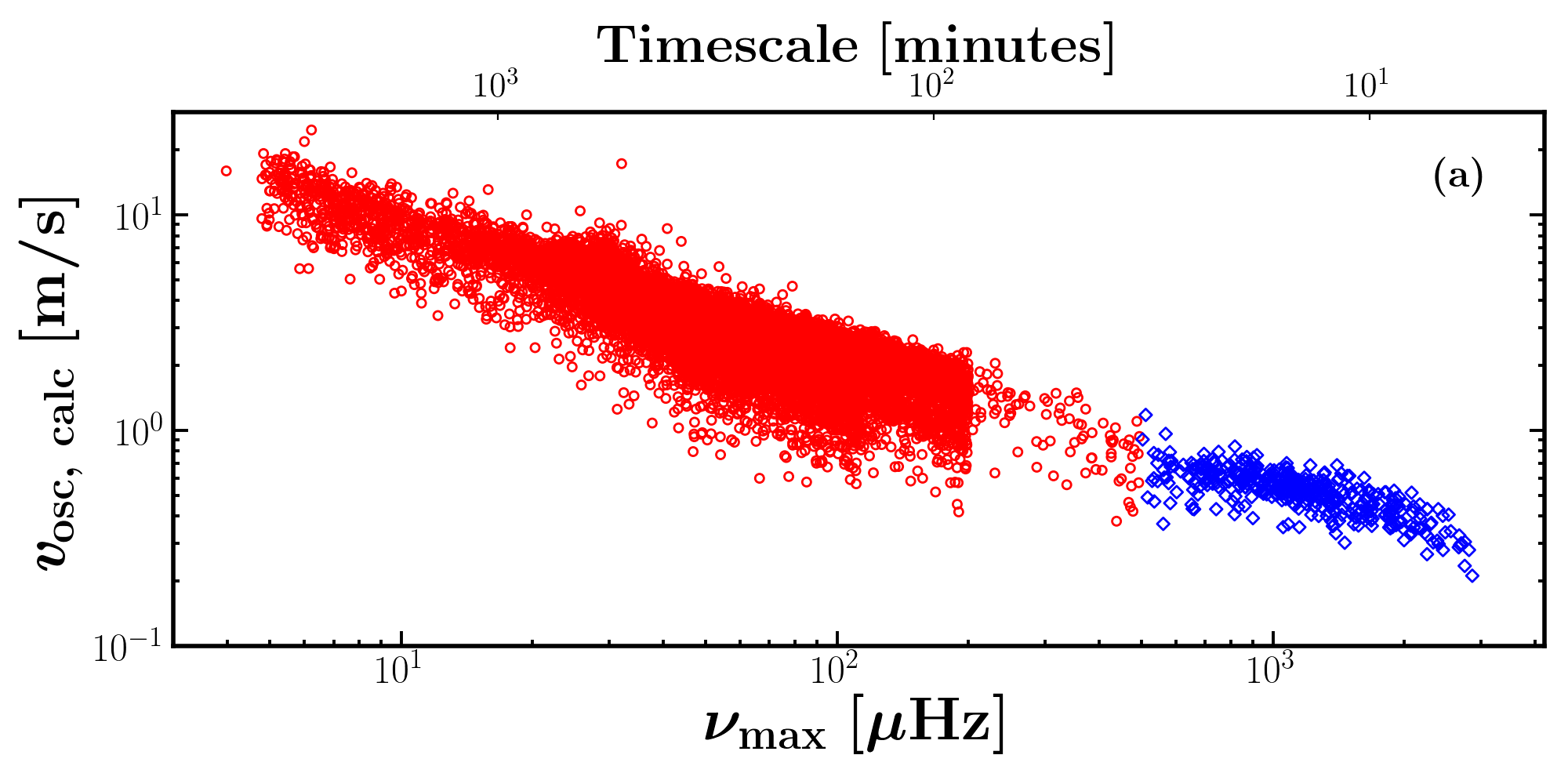}}
\resizebox{\columnwidth}{!}{\includegraphics{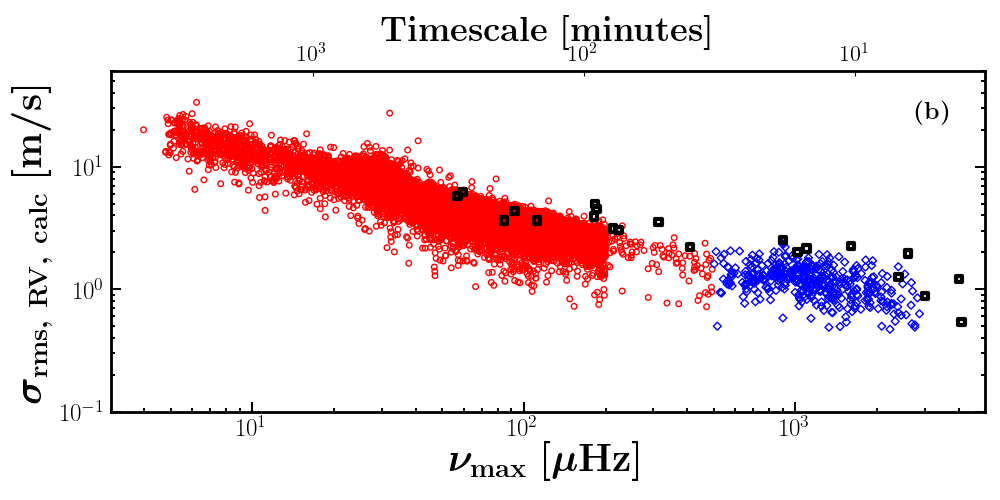}}\\
\caption{\redbf{$\textbf{(a)}$ RV oscillation amplitude and $\textbf{(b)}$ RV jitter due to stellar oscillations. In each panel, the bottom horizontal axis is \numax, while the top horizontal axis is the typical oscillation period (the reciprocal of \numax). The calculated values with different colors are separated with $\numax=500\ \muHz$, used for the subsequent model fitting. An over-density bump at \numax\ $\sim$ 30 \muHz\ arises from red clump stars.}}
\label{fig:figamp}
\end{center}
\end{figure}

The two quantities we seek to predict are the RV oscillation amplitude, \vosc, and RV jitter, \sigmarv. It is important to keep in mind that the granulation background in RV is much lower than in photometry \citep{bedding06}. Therefore, we cannot simply convert the jitter from the photometric time series to its counterpart in the RV time series. Instead, we must first subtract the contributions from granulation and photon noise. This is done most easily by working with the Fourier power spectrum.

\begin{table}
\centering
\caption{Fitted parameters and the uncertainties of Equations \ref{fitfunc1}, \ref{fitfunc2}, and \ref{fitfunc3} for the photometric oscillation amplitude, \amp, RV oscillation amplitude, \vosc, photometric stellar jitter, \sigmaphot, and RV stellar jitter \sigmarv.} 
	\label{tab:fit}
    \resizebox{\columnwidth}{!}{
	\begin{tabular}{rrrrrr} 
        &\multicolumn{4}{c}{Giants, model: $F=F(L, M, T_{\rm eff})$, Equation \ref{fitfunc1}}& \\
        \hline
        \hline
		Parameter & $\alpha\ [\rm m/s]$ & $\beta$ & $\gamma$ & $\delta$ & $\epsilon$\\
 		\hline
		\amp        & $ 7.34\pm0.07$ & $ 0.58\pm0.01$ & $-1.33\pm0.01$ & $-3.50\pm0.04$ & -\\   
		\vosc       & $ 0.31\pm0.01$ & $ 0.60\pm0.01$ & $-1.32\pm0.01$ & $-2.10\pm0.04$ & -\\     
		\sigmaphot  & $11.65\pm0.11$ & $ 0.58\pm0.01$ & $-1.14\pm0.01$ & $-3.33\pm0.03$ & -\\       
        \sigmarv    & $ 0.58\pm0.01$ & $ 0.59\pm0.01$ & $-1.15\pm0.01$ & $-1.55\pm0.03$ & -\\
        \hline
        \hline 
        \\
        &\multicolumn{4}{c}{Giants, model: $F=F(L, T_{\rm eff}, g)$, Equation \ref{fitfunc2}}& \\
        \hline
        \hline
		Parameter & $\alpha\ [\rm m/s]$ & $\beta$ & $\gamma$ & $\delta$ & $\epsilon$ \\
 		\hline
		\amp       &  $ 7.30\pm0.07$&$-0.75\pm0.01$& -&$1.82\pm0.05$&$-1.33\pm0.01$\\   
		\vosc      &  $ 0.31\pm0.01$&$-0.72\pm0.01$& -&$3.20\pm0.05$&$-1.32\pm0.01$\\     
		\sigmaphot &  $11.59\pm0.11$&$-0.56\pm0.01$& -&$1.21\pm0.04$&$-1.14\pm0.01$\\       
        \sigmarv   &  $ 0.58\pm0.01$&$-0.56\pm0.01$& -&$3.04\pm0.04$&$-1.15\pm0.01$\\
        \hline
        \hline 
        \\
        &\multicolumn{4}{c}{Giants, model: $F=F(T_{\rm eff}, g)$, Equation \ref{fitfunc3}}&\\
        \hline
        \hline        
		Parameter & $\alpha\ [\rm m/s]$ & $\beta$ & $\gamma$ & $\delta$ & $\epsilon$\\
		\hline 
		\amp       &  $4.05\pm0.06$&-&-&$-2.15\pm0.06$&$-0.77\pm0.06$\\  
		\vosc      &  $0.13\pm0.01$&-&-&$-0.76\pm0.06$&$-0.63\pm0.06$\\       
		\sigmaphot &  $7.53\pm0.09$&-&-&$-1.59\pm0.05$&$-0.45\pm0.09$\\        
        \sigmarv   &  $0.34\pm0.01$&-&-&$ 0.16\pm0.05$&$-0.45\pm0.09$\\
		\hline
        \hline
        \\
        &\multicolumn{4}{c}{Giants, model: $F=F(L, T_{\rm eff})$, Equation \ref{fitfunc4}}&\\
        \hline
        \hline        
		Parameter  & $\alpha\ [\rm m/s]$ & $\beta$ & $\gamma$ & $\delta$  & $\epsilon$\\
		\hline 
		\amp       & $ 6.15\pm0.13$ & $0.33\pm0.01$ & -&$-0.77\pm0.06$ &-\\  
		\vosc      & $ 0.19\pm0.01$ & $0.37\pm0.01$ & -&$-0.63\pm0.06$ &-\\       
		\sigmaphot & $10.82\pm0.21$ & $0.37\pm0.01$ & -&$-0.45\pm0.09$ &-\\        
        \sigmarv   & $ 0.46\pm0.01$ & $0.39\pm0.01$ & -&$-0.45\pm0.09$ &-\\
		\hline
        \hline        
        \\
        &\multicolumn{4}{c}{Dwarfs\ \&\ subgiants, model: $F=F(L, M, T_{\rm eff})$, Equation \ref{fitfunc1}}&\\
        \hline
        \hline
		Parameter & Model& $\alpha\ [\rm m/s]$ & $\beta$ & $\gamma$ & $\delta$\\
		\hline
		\amp       & $ 5.09\pm0.11$&$0.58\pm0.02$&$-0.77\pm0.06$&$-2.88\pm0.17$ &-\\
		\vosc      & $ 0.30\pm0.01$&$0.50\pm0.02$&$-0.63\pm0.06$&$-0.96\pm0.16$ &-\\
		\sigmaphot & $11.66\pm0.37$&$0.47\pm0.03$&$-0.45\pm0.09$&$-1.43\pm0.20$ &-\\
        \sigmarv   & $ 0.63\pm0.02$&$0.47\pm0.03$&$-0.45\pm0.09$&$0.57\pm0.20$   &-\\  
		\hline
        \hline
        \\       
        &\multicolumn{4}{c}{Dwarfs\ \&\ Subgiants, model: $F=F(L, T_{\rm eff}, g)$, Equation \ref{fitfunc2}}& \\
        \hline
        \hline
		Parameter  & $\alpha\ [\rm m/s]$ & $\beta$ & $\gamma$ & $\delta$ & $\epsilon$\\
		\hline
		\amp       &  $ 5.08\pm0.11$ & $-0.19\pm0.05$ &- & $0.20\pm0.35$ & $-0.77\pm0.06$ \\
		\vosc      &  $ 0.30\pm0.01$ & $-0.13\pm0.05$ &- & $1.57\pm0.35$ & $-0.63\pm0.06$ \\
		\sigmaphot &  $11.64\pm0.37$ & $ 0.01\pm0.07$ &- & $0.39\pm0.47$ & $-0.45\pm0.09$ \\
        \sigmarv   &  $ 0.63\pm0.02$ & $ 0.01\pm0.07$ &- & $2.38\pm0.47$ & $-0.45\pm0.09$  \\  
		\hline
        \hline  
        &\multicolumn{4}{c}{Dwarfs\ \&\ subgiants, model: $F=F(T_{\rm eff}, g)$, Equation \ref{fitfunc3}}& \\
        \hline
        \hline
        Parameter  & $\alpha\ [\rm m/s]$ & $\beta$ & $\gamma$ &$\delta$ & $\epsilon$\\
		\hline
		\amp       & $ 5.09\pm0.11$  &- &- & $-1.05\pm0.15$ & $-0.77\pm0.06$ \\      
		\vosc      & $ 0.30\pm0.01$  &- &- & $ 0.73\pm0.15$ & $-0.63\pm0.06$ \\        
		\sigmaphot & $11.63\pm0.37$  &- &- & $ 0.46\pm0.19$ & $-0.45\pm0.09$ \\
        \sigmarv   & $ 0.63\pm0.02$  &- &- & $ 2.46\pm0.20$ & $-0.45\pm0.09$ \\
		\hline
        \hline 
        \\
        &\multicolumn{4}{c}{Dwarfs\ \&\ subgiants, model: $F=F(L, T_{\rm eff})$, Equation \ref{fitfunc4}}& \\
        \hline
        \hline
        Parameter  & $\alpha\ [\rm m/s]$ & $\beta$ & $\gamma$ & $\delta$  & $\epsilon$\\
		\hline
		\amp       &  $ 5.45\pm0.13$ & $0.41\pm0.02$ &- & $-0.77\pm0.06$ & -\\      
		\vosc      &  $ 0.33\pm0.01$ & $0.34\pm0.02$ &- & $-0.63\pm0.06$ & -\\        
		\sigmaphot &  $12.17\pm0.38$ & $0.36\pm0.02$ &- & $-0.45\pm0.09$ & -\\
        \sigmarv   &  $ 0.66\pm0.02$ & $0.36\pm0.02$ &- & $-0.45\pm0.09$ & -\\
		\hline
        \hline 
        \\  
	\end{tabular}}
\end{table}

First, we \redbf{calculated} the photometric oscillation amplitude, $A_{\lambda}$, \redbf{which was then converted} to the RV amplitude, \vosc.  Specifically, the quantity $A_{\lambda}$ \redbf{was} defined as the oscillation amplitude per radial mode in this manner: 
\begin{equation}
\label{amp}
A_{\lambda} = \frac{\sqrt{\frac{H_{\rm env} \Delta\nu}{\it{c}}}}{\rm {sinc\left(\frac{\pi}{2} \frac{\nu_{max}}{\nu_{Nyq}}\right)}},
\end{equation} 
where, $H_{\rm env}$ is the height of the oscillation power excess in the power spectrum, \Dnu\ is the mean large frequency separation between modes of the same angular degree and consecutive radial orders, $c$ is the effective number of modes per order, adopted as 3.04 \citep{bedding10, stello11}, \numax\ is the frequency of maximum oscillation power, and $\nu_{\rm Nyq}$ is the Nyquist frequency. Note that $\nu_{\rm Nyq}$ is equal to 283~\muHz\ for the \kep\ long-cadence (29.4 minutes) time series and 8333~\muHz\ for the \kep\ short-cadence (58.89 seconds) time series. The attenuation of \redbf{the} oscillation amplitude due to the integration of photons every long- or short-cadence interval \redbf{was} corrected with the sinc function \citep{huber10, murphy12, chaplin14}. 

From the photometric oscillation amplitude $A_{\lambda}$, we \redbf{were} able to obtain the RV amplitude \vosc\ via the relation given by \citet{KB95}:
\begin{equation}
\label{vosc}
\textit{v}_{\rm osc}  = \rm {(A_{\lambda}/20.1 ppm)\ (\lambda/550\ nm)\ (\teff/5777\ K)^2\ [m\ s^{-1}]},
\end{equation}
where \teff\ is the effective temperature, \redbf{and} $\lambda=600\ {\rm nm}$ \redbf{was taken} as a representative wavelength for the broad bandpass of the \kep\ telescope. 

Next, we \redbf{calculated} the photometric jitter \sigmaphot, \redbf{which was then converted} to \sigmarv. Following \citet{kjeldsen92}, the quantity \sigmaphot\ \redbf{was} measured as
\begin{equation}
\label{sigmarms}
\sigma _{\rm rms,\ phot} = \sqrt{\frac{\sigma_{\rm PS} N}{4}},
\end{equation}
where $\sigma_{\rm PS}$ is the mean `noise' level of oscillations (our jitter) in the power spectrum, and $N$ is the number of data points of the time series. In practice, we \redbf{calculated} $\sigma_{\rm PS} \cdot N$ from a power-density spectrum, which is the power spectrum with its power multiplied by the effective observing time \citep{kjeldsen08}. We \redbf{evaluated} the area under the oscillation power excess that can be appropriately approximated with a Gaussian. Thus, we have
\begin{equation}
\label{noise}
\sigma_{\rm PS} \cdot N = \sqrt{\frac{\pi}{4\rm ln2}}H_{\rm env}\it{W},
\end{equation}
where $\it{W}$ is the full-width-at-half-maximum of the oscillation power excess. 

To convert the calculated photometric jitter \sigmaphot\ to the RV jitter \sigmarv, we \redbf{used} Equation \ref{vosc} by replacing \vosc\ and $A_{\lambda}$ with \sigmarv\ and $\sigma _{\rm rms, phot}$, respectively.
Note that we distinguish the \textit{calculated} and \textit{predicted} \sigmarv\ in this work. The former refers to the quantity we derive from Equations \ref{vosc}, \ref{sigmarms}, and \ref{noise}, with the observables $H_{\rm env}$ and $\it{W}$, while the latter refers to the quantity we \redbf{infer} from a fitted model with stellar parameters (see Section \ref{result} for more detail). This naming distinction is also applicable to three other quantities, namely $A_{\lambda}$, \vosc\, and \sigmaphot.

\begin{figure}
\begin{center}
\resizebox{\columnwidth}{!}{\includegraphics{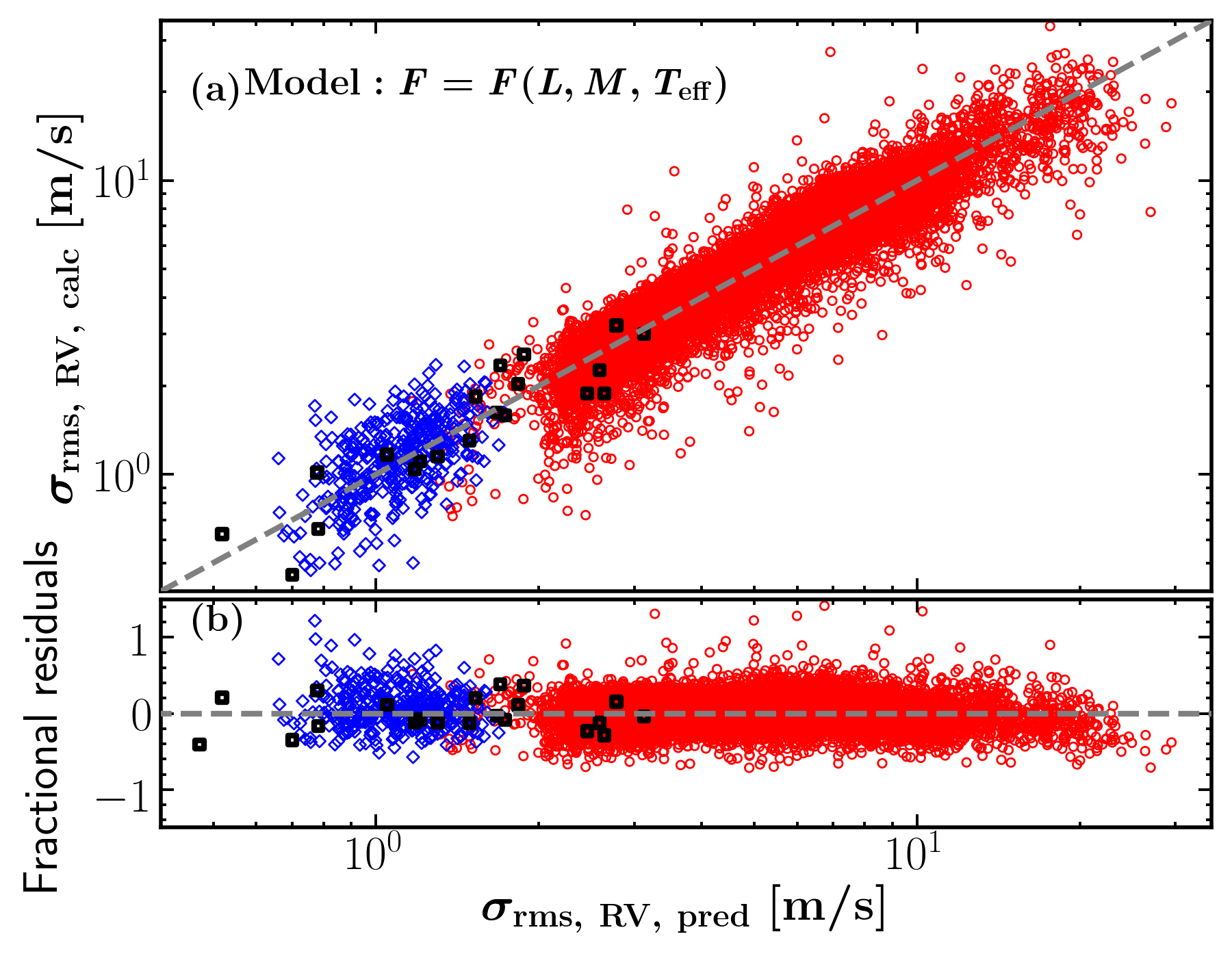}}
\resizebox{\columnwidth}{!}{\includegraphics{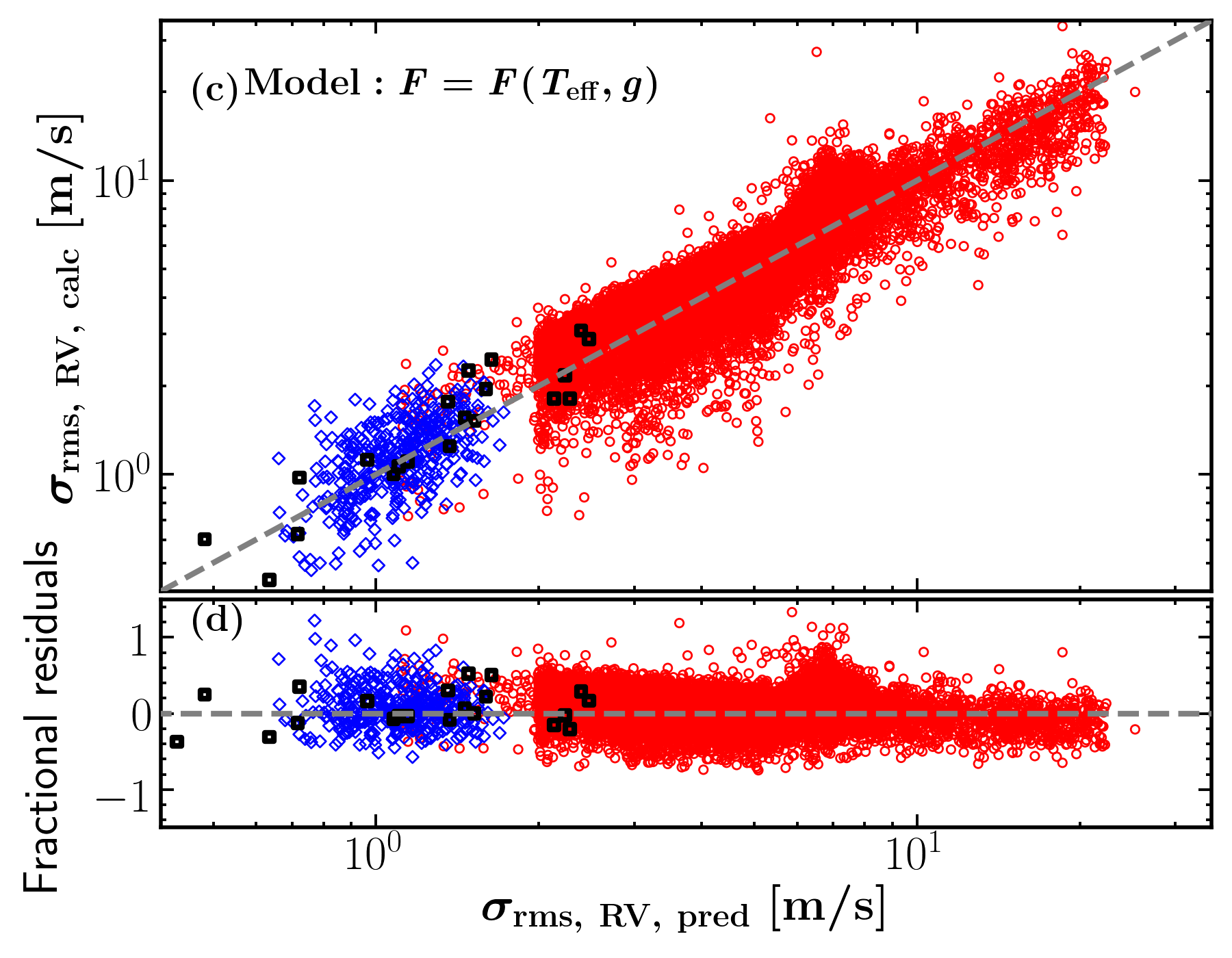}}\\
\resizebox{\columnwidth}{!}{\includegraphics{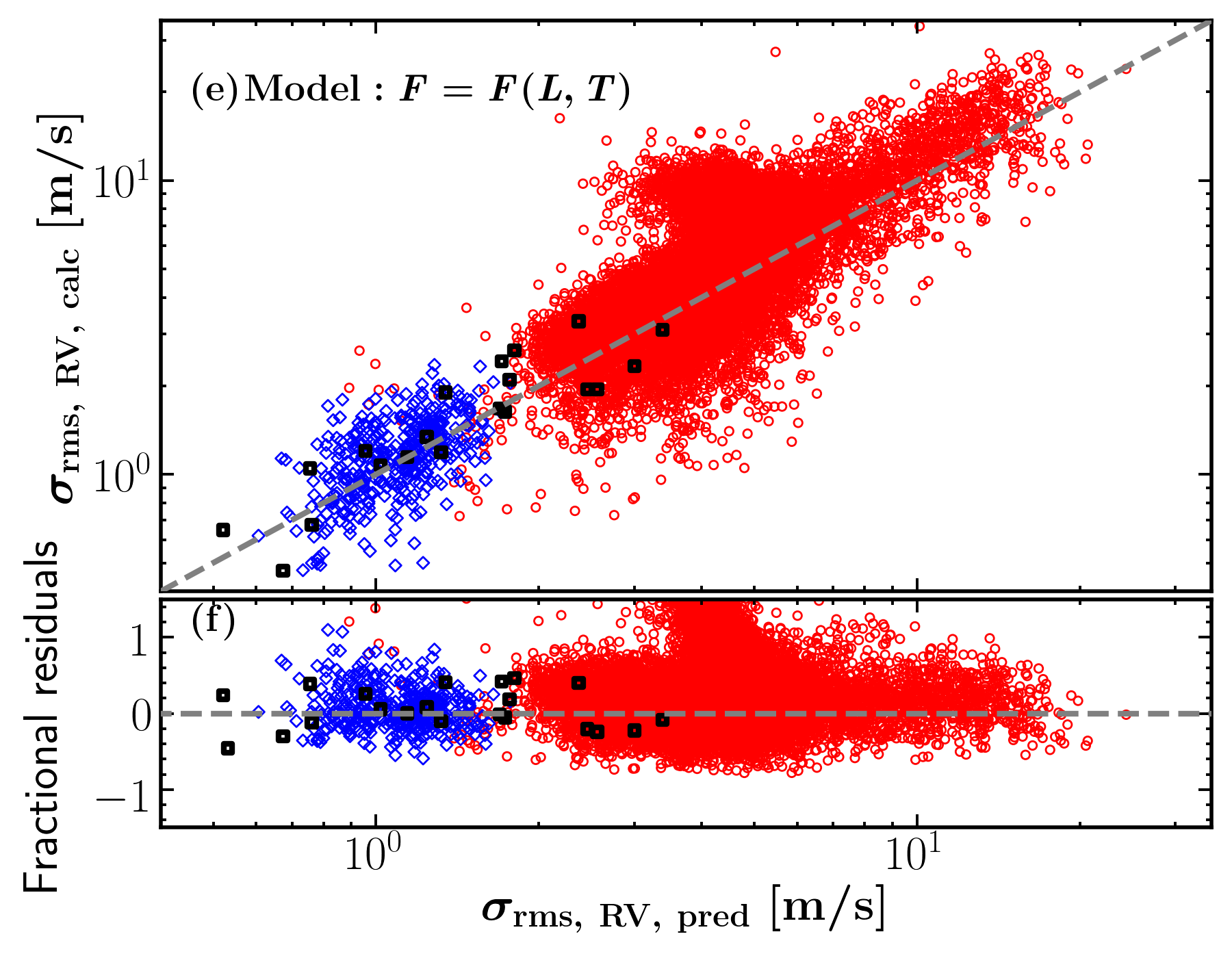}}\\
\caption{\redbf{Comparison of the \textit{calculated} with the \textit{predicted} RV jitter \sigmarv\ (see the text for their definitions) using three models as indicated. Grey dashed lines represent perfect agreement. We separately fitted both dwarfs and subgiants (blue diamonds), and giants (red squares) using a dividing point $\numax = 500\ \muHz$, or equivalently \logg\ $\sim$3.5 dex. The fractional residuals are defined as ($\rm \sigma_{rms, RV, calc}-\sigma_{rms, RV, pred})/\sigma_{rms, RV, pred}$. The bump at \sigmarv\ $\simeq$ 4 m/s\ is caused by red clump stars. Black squares indicate stars with long RV time series from which we computed \sigmarv\ and rescaled it to include contributions only from oscillations (see the text). Here we do not show the comparisons for \amp, \vosc, and \sigmaphot\ given their almost identical features.}}

\label{fig:rmsrv}
\end{center}
\end{figure}

\begin{table*}
\centering
\caption{\redbf{Stellar parameters for 21 bright stars with RV time series observed by ground-based telescopes.}} 
\label{tab:21stars}
\resizebox{\textwidth}{!}{
 \begin{tabular}{rrrrrrrrrr} 
 \hline
 Star           & HD      & \numax  & \sigmarv & $\rm Ref. 1^{a}$   & Luminosity      & Mass            & \logg         &  \teff         & $\rm Ref. 2^{b}$\\
                &         & [\muHz] & [m/s]    &                    & L$_{\odot}$     & M$_{\odot}$     & dex           &  [K]           & \\
 \hline
 $\epsilon$ Tau & 28305   & $56.9$  & 5.80     & \citet{stello17}    & $75.54\pm1.80$ & $2.40\pm0.36$   &$2.67\pm0.08$  &  $4746\pm70$   &\citet{stello17}\\
 46 LMi         & 94264   & $59.4$  & 6.20     & \citet{frandsen18}  & $27.42\pm1.38$ & $1.09\pm0.04$   &$2.674\pm0.013$&  $4690\pm50$   &\citet{frandsen18}\\
 $\beta$ Gem    & 62509   & $84.5$  & 3.64     & \citet{stello17}    & $36.50\pm1.69$ & $1.73\pm0.27$   &$2.84\pm0.08$  &  $4935\pm49$   &\citet{stello17}\\
 $\xi$ Hya      & 100407  & $90$    & 4.37     & \citet{stello04}    & $57.65\pm2.39$ & $2.89\pm0.23$   &$2.883\pm0.032$&  $4984\pm54$   &\citet{bruntt10}\\
 18 Del         & 199665  & $112$   & 3.64     & \citet{stello17}    & $33.52\pm1.77$ & $1.92\pm0.30$   &$2.97\pm0.09$  &  $5076\pm38$   &\citet{stello17}\\
 HD 5608        & 5608    & $181$   & 3.92     & \citet{stello17}    & $12.74\pm0.62$ & $1.32\pm0.21$   &$3.17\pm0.08$  &  $4911\pm51$   &\citet{stello17}\\
 6 Lyn          & 45410   & $183$   & 4.94     & \citet{stello17}    & $13.74\pm0.73$ & $1.37\pm0.22$   &$3.18\pm0.09$  &  $4978\pm18$   &\citet{stello17}\\
 $\gamma$ Cep   & 222404  & $185$   & 4.53     & \citet{stello17}    & $11.17\pm0.16$ & $1.32\pm0.20$   &$3.17\pm0.08$  &  $4764\pm122$  &\citet{stello17}\\
 $\kappa$ CrB   & 142091  & $213$   & 3.13     & \citet{stello17}    & $11.20\pm0.17$ & $1.40\pm0.21$   &$3.24\pm0.08$  &  $4876\pm46$   &\citet{stello17}\\
 HD 210702      & 210702  & $223$   & 3.06     & \citet{stello17}    & $12.33\pm0.52$ & $1.47\pm0.23$   &$3.26\pm0.09$  &  $5000\pm44$   &\citet{stello17}\\
 $\nu$ Ind      & 211998  & $313$   & 3.55     & \citet{bedding06}   & $ 6.28\pm0.23$ & $1.00\pm0.13$   &$3.432\pm0.035$&  $5140\pm80$   &\citet{bruntt10}\\
 $\beta$ Aql    & 188512  & $410$   & 2.22     & \citet{kjeldsen08}  & $ 5.73\pm0.19$ & $1.26\pm0.18$   &$3.525\pm0.036$&  $4986\pm111$  &\citet{bruntt10}\\
 Procyon        & 61421   & $900$   & 2.51     & \citet{bedding10b}  & $ 6.77\pm0.20$ & $1.461\pm0.025$ &$3.976\pm0.016$&  $6494\pm48$   &\citet{bruntt10}\\
 $\beta$ Hyi    & 2151    & $1020$  & 2.01     & \citet{bedding07}   & $ 3.41\pm0.13$ & $1.08\pm0.05$   &$3.955\pm0.018$&  $5840\pm59$   &\citet{bruntt10}\\
 $\alpha$ For   & 20010   & $1100$  & 2.14     & \citet{kjeldsen08}  & $ 4.87\pm0.16$ & $1.53\pm0.18$   &$4.003\pm0.033$&  $6015\pm80$   &\citet{bruntt10}\\
 $\gamma$ Ser   & 168723  & $1600$  & 2.25     & \citet{kjeldsen08}  & $ 3.02\pm0.09$ & $1.30\pm0.15$   &$4.169\pm0.032$&  $6115\pm80$   &\citet{bruntt10}\\
 $\alpha$ Cen A & 128620  & $2400$  & 1.26     & \citet{butler04}    & $ 1.47\pm0.05$ & $1.105\pm0.007$ &$4.307\pm0.005$&  $5746\pm50$   &\citet{bruntt10}\\
 $\gamma$ Pav   & 203608  & $2600$  & 1.96     & \citet{mosser08}    & $ 1.52\pm0.05$ & $1.21\pm0.12$   &$4.397\pm0.022$&  $5990\pm80$   &\citet{bruntt10}\\
 18 Sco         & 146233  & $3000$  & 0.88     & \citet{bazot11}     & $1.058\pm0.028$& $1.02\pm0.03$   &$4.45\pm0.02$  &  $5813\pm21$   &\citet{bazot11}\\
 $\tau$ Cet     & 10700   & $4000$  & 1.21     & \citet{teixeira09}  & $0.47\pm0.02$  & $0.79\pm0.03$   &$4.533\pm0.018$&  $5383\pm47$   &\citet{bruntt10}\\
 $\alpha$ Cen B & 128621  & $4100$  & 0.54     & \citet{kjeldsen05}  & $0.47\pm0.02$  & $0.934\pm0.006$ &$4.538\pm0.008$&  $5140\pm56$   &\citet{bruntt10}\\       
\hline 
\end{tabular}}
\raggedright $\rm Ref. 1^{a}$: References for \numax\ and time series used to calculate \sigmarv\ in this work. \\
$\rm Ref. 2^{b}$: References for stellar parameters, luminosities, masses, \logg, and \teff.
\end{table*}

Thus, to calculate $\sigma _{\rm rms, RV}$ and $\it{v}_{\rm osc}$, we need to know $H_{\rm env}$, $\it{W}$, \numax, and \Dnu\ for individual stars.  We \redbf{adopted} the estimates of these global oscillation parameters from \citet{huber11b} and \citet{yu18}. \citet{huber11b} measured these parameters for dwarfs and subgiants using short-cadence \kep\ time series. \citet{yu18} determined these parameters for red giants with a homogeneous analysis of the full-length end-of-mission \kep\ long-cadence data set, using the same analysis pipeline \citep{huber09}.

\section{Predicting RV jitter from stellar parameters}
\label{result}
Figure \ref{fig:figamp}a shows the calculated RV oscillation amplitude, $\it{v}_{\rm osc}$, for dwarfs, subgiants, and giants, while Figure \ref{fig:figamp}b shows the calculated RV jitter \sigmarv. This can be used to predict the RV jitter if \numax\ is known. Black squares mark the measured \sigmarv\ from published RV time series for (ordered by increasing \numax) $\epsilon$ Tau \citep{stello17}, 46 LMi \citep{frandsen18}, $\beta$ Gem \citep{stello17}, $\xi$ Hya \citep{stello04}, 18 Del, HD 5608, 6 Lyn, $\gamma$ Cep, $\kappa$ CrB, HD 210702 \citep{stello17}, $\nu$ Ind \citep{bedding06}, $\beta$ Aql \citep{kjeldsen08}, Procyon \citep{bedding10b}, $\beta$ Hyi \citep{bedding07}, $\alpha$ For, $\gamma$ Ser \citep{kjeldsen08}, $\alpha$ Cen A \citep{butler04}, $\gamma$ Pav \citep{mosser08},  18 Sco \citep{bazot11},  $\tau$ Cet \citep{teixeira09},  $\alpha$ Cen B \citep{kjeldsen05}. The estimates of \numax\ were adopted from the corresponding literature \redbf{and are given in Table \ref{tab:21stars}}. We can see that the measured \sigmarv\ values are slightly higher than those of \kep\ target stars at a similar \numax. This is due to the additional contributions from granulation at various timescales, as well as from instrumental and photon noise, in particular for dwarfs. We thus suggest to multiply the observed jitter \sigmarv\ due to the oscillations, as done in this work, by a correction factor to approximate the total RV jitter containing oscillations and granulations \redbf{(see the subsequent text).}

Our ultimate goal is to predict \sigmarv\ in terms of fundamental stellar properties. For this, we \redbf{used} four simple models. The first model is
\begin{equation}
\label{fitfunc1}
F = \alpha \left(\frac{L}{L_{\odot}}\right)^\beta \left(\frac{M}{M_{\odot}}\right)^\gamma \left(\frac{T_{\rm{eff}}}{T_{\rm{eff}\odot}}\right)^\delta, 
\end{equation}
where, $L, M$, and $T_{\rm eff}$ are luminosity, mass, and effective temperature, respectively, and $F$ is the quantity that we seek to fit, namely one of \sigmarv, \sigmaphot, $A_{\lambda}$, and \vosc, by adjusting the free parameters, $\alpha$, $\beta$, $\gamma$, and $\delta$. For typical exoplanet host stars, masses may not always be available, we therefore also \redbf{fitted} a second model by substituting the mass, $M$, with surface gravity, $g$,
\begin{equation}
\label{fitfunc2}
F = \alpha \left(\frac{L}{L_{\odot}}\right)^\beta  \left(\frac{T_{\rm{eff}}}{T_{\rm{eff}\odot}}\right)^\delta \left(\frac{g}{g_{\odot}}\right)^\epsilon, 
\end{equation}
where $\epsilon$ is a free parameter. In addition, we \redbf{fitted} the following two models to cater for cases where only \teff\ and $g$, or L and \teff\, are known:
\begin{equation}
\label{fitfunc3}
F = \alpha  \left(\frac{T_{\rm{eff}}}{T_{\rm{eff}\odot}}\right)^\delta
\left(\frac{g}{g_{\odot}}\right)^\epsilon, 
\end{equation}
and
\begin{equation}
\label{fitfunc4}
F = \alpha \left(\frac{L}{L_{\odot}}\right)^\beta \left(\frac{T_{\rm{eff}}}{T_{\rm{eff}\odot}}\right)^\delta. 
\end{equation}
\redbf{The last model is analogous to the one used by \citet{wright05}, who linked the magnitude of RV jitter with $B-V$ color and absolute magnitude of a star.} In the four models, we \redbf{introduced} the coefficient $\alpha$ which allows for our models to not have to pass through the solar reference point. We \redbf{included} luminosity in the models, given that the \textit{Gaia} mission has provided precise parallaxes \citep{lindegren18} and thus luminosities for a large number of stars observed by the \kep\ telescope \citep{berger18,Fulton2018}.

To implement the fit, we used the non-linear least-square minimization code, $LMFIT$, with the Levenberg-Marquardt algorithm \citep{newville16}. We fitted separately giants and dwarfs using $\numax\ =\ 500\ \muHz$, or equivalently \logg\ $\sim3.5$ dex as the dividing point. We calculated luminosities, masses, and surface gravities for the stars in \citet{huber11b}, using the well-known seismic scaling relations \citep{ulrich86,KB95}. For red giants, we took the stellar parameters from \citet{yu18}, which are based on the same relations.  Effective temperatures used in this work were taken from \citet{mathur17}. 

Figure \ref{fig:rmsrv} shows the comparison between the calculated and predicted \sigmarv\ (See Section \ref{method} for the definitions). We can see from Figures \ref{fig:rmsrv}a and \ref{fig:rmsrv}b that luminosity, mass, and temperature can be used to make quite good predictions of the RV jitter \sigmarv\ for both dwarfs and giants. The comparison returns a median fractional residual of 4.4\% with a scatter of 17.9\% for dwarfs, and a median fractional residual of 3.3\% with a scatter of 27.1\% for giants. \redbf{To test the model, we computed \sigmarv\ for 21 stars, as listed in Table~\ref{tab:21stars}, from their real RV time series. Note that the predicted RV jitter are only from oscillations. Thus, we removed granulation contributions from the computed \sigmarv\ for the 21 stars by dividing a correction factor of 1.9. The correction factor was taken to be the median ratio between the measured \sigmarv\ from RV time series, and the predicted \sigmarv, using the model of Equation \ref{fitfunc1} with  L, M, and \teff\ from Table \ref{tab:21stars}. The agreement as shown in black squares is very good, with an offset of $-1.5$\% and a scatter of 22.1\% in the fractional residuals.} The model of Equation \ref{fitfunc2} gives the same fit quality with that of Equation \ref{fitfunc1}, for which the comparison is not shown here. 

Figures \ref{fig:rmsrv}c and \ref{fig:rmsrv}d show that a combination of surface gravity and effective temperature is also capable of making reasonable predictions of \sigmarv, with precisions of 22.6\% for dwarfs and 27.1\% for giants. \redbf{A correction factor of 2.0 for this model is recommended.} In the case where only luminosity and effective temperature are available, we still get a useful prediction of \sigmarv\ for dwarfs and subgiants (47.8\% precision) and giants (27.5\% precision), as shown in \ref{fig:rmsrv}e and \ref{fig:rmsrv}f. \redbf{We suggest a correction factor of 1.9 for this model.} The prominent feature present at \sigmarv\ $\simeq\ 4\ \rm m/s$ is caused by red clump stars that have globally smaller masses than red-giant-branch stars at similar \sigmarv. We do not show the comparison figures for the photometric amplitude \amp, RV oscillation \vosc, and photometric stellar jitter \sigmaphot, because they exhibit similar properties to these of \sigmarv. We provide all the fitted parameter values and their standard deviations in Table \ref{tab:fit}. 

\begin{figure}
\begin{center}
\resizebox{\columnwidth}{!}{\includegraphics{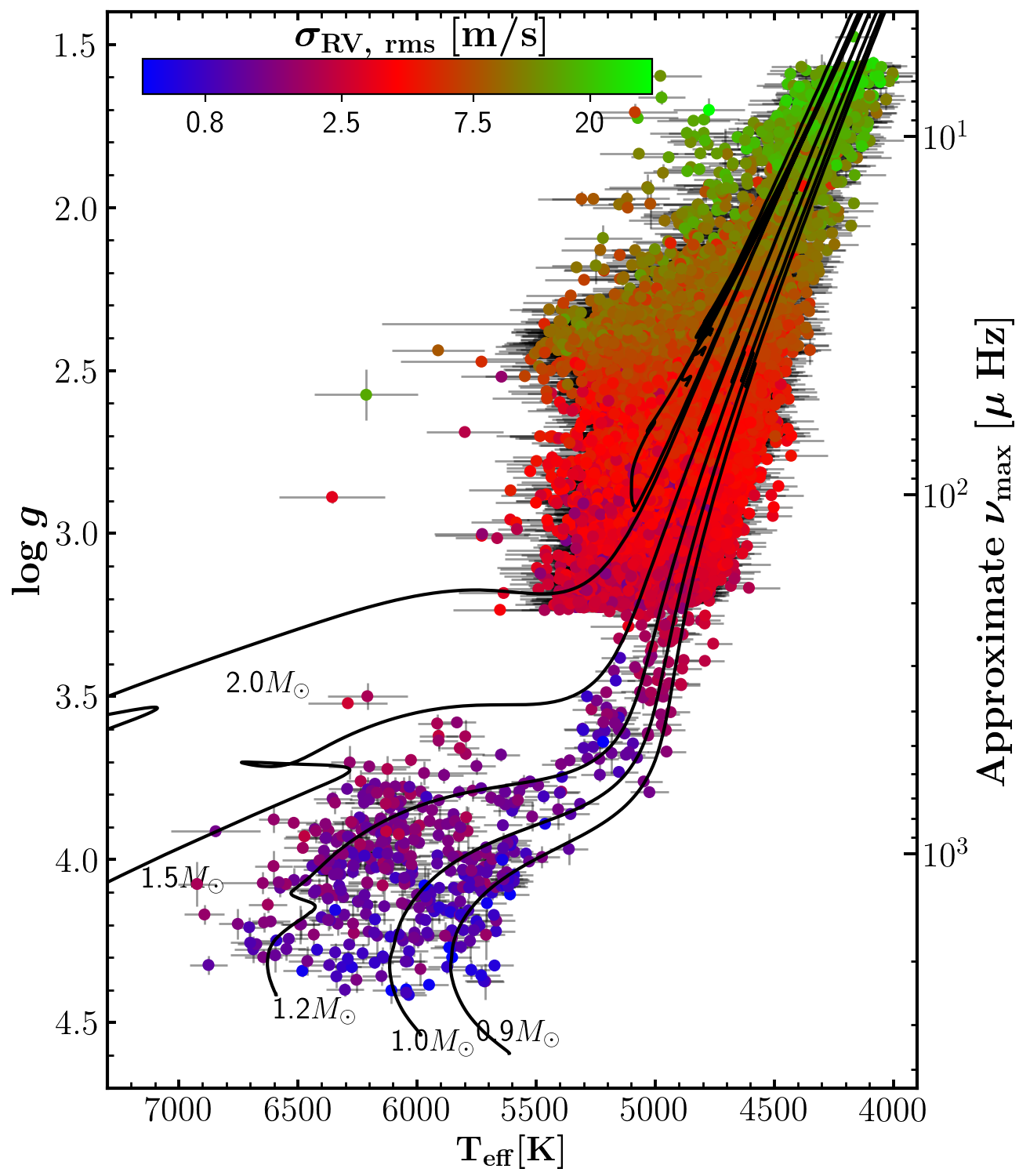}}\\
\caption{$\rm log$ $g$ vs. \teff\ diagram color-coded by the RV jitter \sigmarv. Approximate \numax\ is labeled in the right vertical axis. The solid lines show evolutionary tracks from PARSEC \citep{bressan12}, with the masses from 0.8 to 2.0 $\rm M_{\odot}$ and the metallicy [Fe/H]\ =\ -0.096 equal to the median value of the whole sample.} 
\label{fig:hr}
\end{center}
\end{figure}

Figure \ref{fig:hr} shows the H-R diagram of \kep\ targets, color-coded by the RV jitter \sigmarv. We observe a cutoff of star number density at $\numax\ =~200~\muHz$ due to the transition from short cadence to long cadence. Typically, the RV jitter is at the level of $\sim$~0.5~m/s in dwarfs, $\sim$~1.5~m/s in subgiants, $\sim$~4~m/s in low-luminosity red giants (\numax\ close to 100~\muHz) , $\sim$~7~m/s in red clump stars (\numax\ close to 40~\muHz), and $\sim$~15~m/s in high-luminosity red giants (\numax\ close to 10~\muHz). Encouragingly, these values are consistent with observed jitter values for stars in similar evolutionary states \citep{johnson10, jones13, wittenmyer16, wittenmyer17}.

\section{Conclusions}
We calculated the RV jitter \sigmarv\ due to stellar oscillations using the global oscillation parameters, the height $H_{\rm env}$ and width $W$ of oscillation power excess, measured with \kep\ data. We then predicted the RV jitter in terms of stellar parameters for both dwarfs and giants. Using four sets of stellar parameters, we obtained the following precisions (\redbf{i.e., standard deviations of fractional residuals}):

\begin{enumerate}
\item \ $L,\ M,\ T_{\rm eff}$: 17.9\% for dwarfs and subgiants, 27.1\% for giants.
\item \ $L,\ T,\ g$: 17.9\% for dwarfs and subgiants, 27.1\% for giants.
\item \ $T,\ g$: 22.6\% for dwarfs and subgiants, 27.1\% for giants.
\item \ $L,\ T$: 47.8\% for dwarfs and subgiants, 27.5\% for giants.
\end{enumerate}

A comparison between our calculated RV jitter \sigmarv\ and those directly computed from RV time series indicates that the predicted \sigmarv\ is globally smaller than observed in RV data. This is due to the observed \sigmarv\ values including the extra contributions from granulation, as well as photon noise and instrumental noise. We stress that the RV jitter predicted from this work are only from stellar oscillations, representing the lower limit. A correction factor is suggested to be applied to our predicted \sigmarv, so as to approximate the whole RV jitter including both oscillations and granulation. \redbf{By calibrating on long RV time series, we recommend to increase the estimates by using a factor of 1.9 when using the models of Equation \ref{fitfunc1} and \ref{fitfunc2}, and factors of 2.0 and 1.9 when using the models of Equation \ref{fitfunc3} and \ref{fitfunc4}, respectively.} 

The predicted RV jitter \sigmarv\ can provide guidance to the follow-up spectroscopic observations for the exoplanets to be found by transit surveys, such as the \tess\ and \plato\ space missions. They can also be used to set a prior for the jitter term as a component when modeling Keplerian orbits \citep[e.g.][]{eastman13,fulton18}. We provide publicly available code to estimate the RV jitter \sigmarv.

\section*{Acknowledgements}
We gratefully acknowledge the entire \kep\ team and everyone involved in the \kep\ mission for making this paper possible. Funding for the \kep\ Mission is provided by NASA's Science Mission Directorate. D.H. acknowledges support by the National Aeronautics and Space Administration under Grant NNX14AB92G issued through the Kepler Participating Scientist Program. D.S. is the recipient of an Australian Research Council Future Fellowship (project number FT1400147). 

\bibliographystyle{mnras}
\bibliography{rvamp/references}

\end{document}